\documentclass[twocolumn,amsmath,showpacs,amsfonts,aps,prc,floatfix]{revtex4}
\usepackage{graphicx}
\usepackage{bm}

\begin{document}

\title{A lattice based equation of state and  $\phi$ meson production in $\sqrt{s }$=6-200 GeV  Pb+Pb and Au+Au collisions }
\author{Victor Roy}
\email[E-mail:]{victor@veccal.ernet.in} 
\author{A. K. Chaudhuri}
\email[E-mail:]{akc@veccal.ernet.in}

\affiliation{Variable Energy Cyclotron Centre, 1/AF, Bidhan Nagar, 
Kolkata 700~064, India}

\begin{abstract}

In a boost-invariant hydrodynamic model , we have analyzed $\phi$ meson production in Pb+Pb and Au+Au collisions, in the centre of mass energy range $\sqrt{s }$=
6-200 GeV. Hydrodynamic evolution is governed by a lattice based equation of state with a confinement-deconfinement {\em cross over} at $T_{co}$=196 MeV.
We also look for the threshold energy above which the confined matter undergoes a deconfinement transition. Analysis indicate that above a threshold energy, ${\sqrt{s }}_{th}$=$13.65 \pm 3.06$ GeV, the fluid medium crosses over from a confined phase to a deconfined phase.  
 \end{abstract}


\date{\today}  

\maketitle


\section{Introduction}
\label{intro}
 
Lattice simulations \cite{lattice,Cheng:2007jq} of Quantum Chromodynamics predict that nuclear matter, at extreme conditions, e.g. high temperature or pressure, can undergo a confinement-deconfinement cross-over transition.   The deconfined matter, comprising quarks and gluons, even at very high temperature, is not totally free from interaction and one calls the matter strongly interacting Quark Gluon Plasma (sQGP).
Heavy ion collisions at relativistic energies provides a convenient method to
produce   and study properties QCD matter at extreme conditions. Recent
experiments in Au+Au collisions at RHIC  \cite{BRAHMSwhitepaper,PHOBOSwhitepaper,PHENIXwhitepaper,STARwhitepaper}, produced convincing evidences that in central and mid-central Au+Au collisions, a hot, dense, strongly interacting,  collective QCD matter is created. Whether the matter can be characterized as the lattice QCD \cite{lattice,Cheng:2007jq} predicted Quark-Gluon-Plasma (QGP) or not,   is still a question of debate.

For long, strangeness enhancement is considered as a signature of QGP formation  \cite{Koch:1986ud}. In QGP environment, $gg\rightarrow s\bar{s}$ is abundant.
If not annihilated before hadronisation, early produced strange and anti-strange quarks will coalesce in to strange hadrons and 
compared to elementary pp collisions, strange particle production will be enhanced.  
However, strangeness enhancement could also be obtained in a purely hadronic scenario, mainly due to 'volume effect'  \cite{Rafelski:1980gk,Cleymans:1998yb,Hamieh:2000tk,Tounsi:2002nd}. Strangeness production in small volume elementary pp collisions can be 'canonically' suppressed due to 'strict' strangeness conservation \cite{Rafelski:1980gk,Cleymans:1998yb,Hamieh:2000tk,Tounsi:2002nd}.
In bigger volume AA collisions, locally, strangeness conservation condition can be relaxed to produce strange particles. In the language of statistical mechanics, while canonical ensemble is applicable in pp collision, grand canonical ensemble is more appropriate in heavy ion collisions. Additionally, strange particle phase space appears to be undersaturated in elementary pp or peripheral heavy ion collisions \cite{Becattini:2003wp,Becattini:2005xt}. The deviation from strangeness equilibrium is parameterized by strangeness undersaturation factor $\gamma_S$. 
In a statistical model, fits to particle multiplicities are unsatisfactory, unless
$\gamma_S$ is not accounted for. $\gamma_S$ shows an increasing trend from AGS to RHIC energy. At RHIC, $\gamma_S \sim 1$. A possible interpretation is that the total amount of strangeness available at the hadronisation is determined in the pre-hadronic stage.  The change in $\gamma_S$ between pp and A+A collisions then reflect the difference between initial conditions of the respective systems.   

 Recently STAR collaboration published their measurements of $\phi$ production in Au+Au and Cu+Cu collisions at RHIC energy ($\sqrt{s }$=200 GeV) \cite{Abelev:2007rw,:2008fd,Abelev:2008zk}. Both in Au+Au and Cu+Cu collisions, $\phi$ meson production is enhanced. The enhancement factor is $\sim$ 2. $\phi(s\bar{s})$ mesons are hidden strange particles, unaffected by the 'canonical' suppression. STAR data on $\phi$ meson production is consistent with models with recombination of thermal strange quarks, indicating that in Au+Au collisions a robust partonic system in created \cite{Hwa:2006vb}. Interestingly, strangeness enhancement is observed even at much lower energy.  At SPS,   NA49 collaboration measured $\phi$ meson production in 20A, 30A, 40A, 80A and 158A GeV Pb+Pb collisions \cite{Alt:2008iv}. 
 $\phi$ meson production is enhanced also at SPS energy. Enhancement factor, $\sim$3-4, is even larger than that in RHIC collisions. One naturally wonders about the sources of $\phi$ meson enhancement in low SPS energy collisions. Is it due to   formation of a deconfined medium like in Au+Au collisions at RHIC?  NA49 collaboration analyzed the data in the statistical hadronisation model \cite{Becattini:2003wp}. Statistical hadronisation model, including the  strangeness undersaturation factor $\gamma_S$, provides a good description of the data. Apparently, as in Au+Au collisions at RHIC, initial partonic content of the system also  drive the $\phi$ meson production
in Pb+Pb collisions  at SPS.
 
Statistical hadronisation model is not a dynamic model. 
A statistical model can not comment whether or not the initial system is in a deconfined state.
Relativistic hydrodynamics is the only dynamical model, which can, within certain limitations, comment on the initial condition of the fluid. It is assumed that in the collision a fireball is produced. Constituents of the fireball collide frequently to establish local thermal equilibrium sufficiently fast and after a certain (proper) time $\tau_i$, hydrodynamics become applicable. If the macroscopic properties of the fluid e.g. energy density, pressure, velocity etc. are known at the equilibration time $\tau_i$, the relativistic hydrodynamic equations (i.e. energy-momentum conservation equations) can be solved to give the space-time evolution of the fireball till a given freeze-out condition such that interactions between the constituents are too weak to continue the evolution. 
Using suitable algorithm (e.g. Cooper-Frye) information at the freeze-out can be converted into particle spectra and can be directly compared with the experimental data. Thus, hydrodynamics, in an indirect way, can characterize the initial condition of the medium produced in heavy ion collisions. Hydrodynamics equations are closed only with an equation of state (EOS) and one can investigate the possibility of phase transition in the medium.  

In the present paper, in an ideal hydrodynamic model, we have analyzed the NA49
data \cite{Alt:2008iv} on $\phi$ meson production in 20-158 A GeV central Pb+Pb collisions and the STAR data \cite{Abelev:2007rw,:2008fd,Abelev:2008zk} on $\phi$ meson production  in $\sqrt{s }$=62, 130 and 200 GeV central Au+Au collisions, effectively scanning the c.m. energy range $\sqrt{s }$=6-200 GeV. 
An important ingredient of the model is a lattice based EOS with a confinement-deconfinement cross over transition at $T_{co}$=196 MeV. Our analysis indicate that ideal hydrodynamics,
with the lattice based EOS, reasonably well explain the $\phi$ meson production in the energy range,
$\sqrt{s}$=6-200 GeV. 
Analysis also indicate that   above a
threshold energy $\sqrt{s}_{th}=13.65 \pm 3.06$ GeV,  the medium crosses over from a confined phase to a deconfined phase. It was also indicated that the transition is rather sharp, occuring over a narrow energy range
$\Delta\sqrt{s}_{th}=11.28 \pm 5.93$ GeV.
  
The paper is organized as follows: in section \ref{sec2}, we briefly describe the 
construction of a lattice based equation of state. Initial conditions used to compute fluid evolution etc., are also discussed in section \ref{sec2}. Experimental data e.g. $\phi$ meson's multiplicity, $p_T$-spectra, mean $p_T$ are analyzed in section \ref{sec3}. Summary and conclusions are given in section \ref{sec4}.

\section{Hydrodynamical equations, equation of state and initial conditions}
\label{sec2}
 
Assuming boost-invariance, we have solved the energy-momentum conservation equation, $\partial_\mu T^{\mu\nu}=0$ in $(\tau=\sqrt{t^2-z^2},x,y,\eta_s=\frac{1}{2}\ln\frac{t+z}{t-z})$ coordinates,
using the code "`AZHYDRO", details of which can be found in \cite{QGP3}.
As mentioned earlier, we have used a lattice based equation of state. 
 Equation of state (EOS) is  one of the most important inputs of a hydrodynamic model. Through this input macroscopic hydrodynamic models make contact with the microscopic world.  
Most of the hydrodynamical calculations are performed with EOS with a 1st order phase transition. However, lattice simulations \cite{Cheng:2007jq} indicate that confinement to deconfinement transition is a cross over, rather than a 1st or 2nd order phase transition. It is then essential that hydrodynamic simulations are done with EOS with cross-over transition rather than with EOS with 1st or 2nd order transition.
Huovinen    \cite{Huovinen:2005gy} reported an 'ideal' hydrodynamic simulation with a cross-over phase transition. He concluded that an EOS with 1st order phase transition better explain the experimental data (e.g. elliptic flow of proton or antiproton) than an EOS with cross over transition.

\begin{figure}[t]
\vspace{0.3cm} 
\center
 \resizebox{0.35\textwidth}{!}{%
  \includegraphics{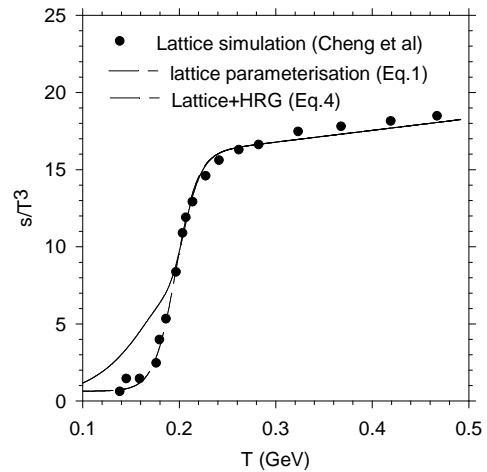}
}
\caption{Black circles are lattice simulation \cite{Cheng:2007jq} for entropy density. The dash line is a parameterisation of lattice simulation by Eq.\ref{eq1}. The solid line is the entropy density in the lattice+HRG model (see text). }\label{F1}
\end{figure}

Presently, we   have used an EOS based on 
recent lattice simulation of Cheng et al. \cite{Cheng:2007jq}. 
 In Fig.\ref{F1},  simulation results \cite{Cheng:2007jq} for the  entropy density is
  shown.    The dotted line in Fig.\ref{F1} is a parameterisation of the entropy density. 
  
\begin{equation}\label{eq1}
\frac{s}{T^3}=\alpha+[\beta+\gamma T][1+tanh\frac{T-T_c}{\Delta T}],
\end{equation}

\noindent with $\alpha$=0.64, $\beta$=6.93, $\gamma$=0.55, $T_c$=196 MeV,
$\Delta T=0.1T_c$.
From the parametric form of the entropy density, pressure and energy density can be obtained using the thermodynamic relations,

\begin{eqnarray}  
  p(T)&=&\int_0^T s(T) dT \label{eq2a} \\
  \varepsilon(T)&=&Ts -p \label{eq2b}.
  \end{eqnarray}
  
\begin{figure}[t]
\center
 \resizebox{0.35\textwidth}{!}{%
  \includegraphics{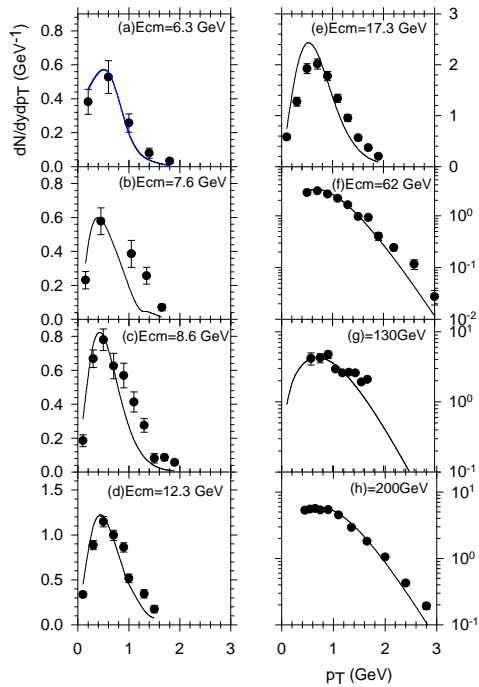}
}
\caption{ In panels (a)-(e), transverse momentum spectra for $\phi$ mesons
in Pb+Pb collisions at $\sqrt{s }$=6.3, 7.6, 8.8, 12.3 and 17.3 GeV
are shown. 
 STAR measurements of $p_T$ spectra for $\phi$ in $\sqrt{s }$=62,130 and 200 GeV Au+Au collisions are shown in panels (f)-(h).  The     lines are respectively are ideal hydrodynamic fit to the $\phi$ spectra. The blue line in panel (a), is a fit obtained to the data when confinement-deconfinement transition occur at $T_{co}$=160 MeV.}  \label{F2}
\end{figure} 


We complement the lattice simulated EOS \cite{Cheng:2007jq} by a
  hadronic resonance gas (HRG) EOS comprising all the resonances below mass 2.5 GeV. In Fig.\ref{F1}, the solid line is the
   entropy density of the "`lattice +HRG"' EOS. The entropy density of the complete EOS is obtained as,
     
   \begin{equation}
   s=0.5[1+tanh(x)]s_{HRG} + 0.5 [1-tanh(x)]s_{LATTICE}
   \end{equation}
   
\noindent   with $x=\frac{T-T_c}{\Delta T}$.   Compared to lattice simulation, entropy density in HRG drops slowly at low temperature.
It is consistent with the observation that at low temperature, trace anomaly,
$\frac{\varepsilon-3p}{T^4}$ drops faster in lattice simulation than in a HRG model \cite{Cheng:2007jq}. It is difficult to resolve whether the discrepancy between lattice simulations at low temperature and HRG model is due to failure of HRG model at lower temperature
or due to the difficulty in resolving low energy hadron spectrum on a  rather coarse lattice \cite{Cheng:2007jq}.

   

We note that, lattice simulations \cite{Cheng:2007jq} were performed with zero baryon density $\mu_B$=0. $\mu_B$=0 is only approximately valid in RHIC energy collisions. In lower SPS energy collisions, there can be considerable stopping and  $\mu_B$ can be appreciably different from zero.   However, lattice simulations with finite baryon density have yet not reached the state of application. We therefore continue to use the $\mu_B$=0 lattice based EOS,  even at low SPS energy collisions. Later, we will comment on the possible 
effects finite baryon density EOS can have on our analysis.
 
\begin{figure}[t]
\center
 \resizebox{0.35\textwidth}{!}{%
  \includegraphics{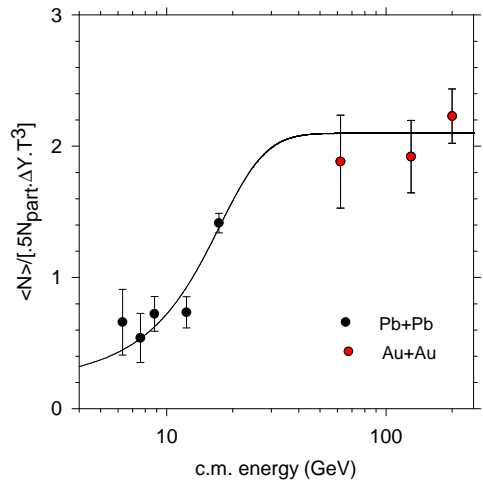}
}
\caption{ Ratio of the mean $\phi$ meson multiplicity per participant
over cube of the initial temperature, normalised by the participant number ($N_{part}$) and rapidity width ($\Delta Y$),
in   Pb+Pb and   Au+Au collisions as a function of collisions energy. The solid line is a fit to the ratio with the form, $\alpha [1+tanh\frac{\sqrt{s }-{\sqrt{s }}_{th}}{\Delta}]$. }    \label{F3}
\end{figure}   
 
We assume that at the initial time $\tau_i$=0.6 fm, the initial energy density
of the fluid is  distributed as \cite{QGP3}

\begin{equation} \label{eq6}
\varepsilon({\bf b},x,y)=\varepsilon_0[0.75 N_{part}({\bf b},x,y) +0.25 N_{coll}({\bf b},x,y)],
\end{equation}

\noindent
where b is the impact parameter of the collision. $N_{part}$ and $N_{coll}$ are the transverse profile for the average participant and collision number respectively. $N_{part}$ and $N_{coll}$ are calculated in a Glauber model, with nuclear density,

\begin{equation}
\rho(r)=\frac{\rho_0}{1.+e^{(r-R)/a}},
\end{equation} 

\noindent with R=6.624 (6.37) fm and a=0.549 (0.54) fm for Pb (Au) nucleus. $\rho_0$ is determined from the normalization condition, $\int \rho(r)d^3r$ = A (208 for Pb, 197 for Au). Glauber model calculation also require the total inelastic NN cross section $\sigma_{inel}$.
NN cross-section varies slowly with energy. Throughout the energy range, $\sqrt{s }$=6.3-17 GeV, we assume $\sigma_{inel}$=32 mb. For $\sqrt{s }$=62, 130 and 200 GeV Au+Au collisions, we have used, $\sigma_{inel}$=36, 40  and 42 mb respectively. We also assume that at the initial  
time $\tau_i$, the transverse fluid velocity is zero. 
 $\varepsilon_0$ in Eq.\ref{eq6} corresponds to  central energy density in zero impact parameter collisions.  We fix it to reproduce the experimental $p_T$ spectra of $\phi$ mesons from a freeze-out surface at $T_F$=150 MeV \cite{note1}.

\begin{table*}[t]
\caption{ \label{table1} Details of the data analysed are given. The Pb+Pb data are from the NA49 experiment  \cite{Alt:2008iv}. The Au+Au data are from the STAR experiment \cite{:2008fd}.
$b$ is the impact parameter corresponding to the centrality cut. The fitted values of the central energy density, $\varepsilon_0$ and corresponding $\chi^2/N$ of the fit,
ratio of experimental $\phi$ multiplicity over the hydrodynamic predictions $<N>^{ex}/<N>^{TH}$ and the ratio of experimental $\phi$ meson mean $p_T$ over the predicted mean $p_T$, $<p_T>^{ex}/<p_T>^{TH}$ are shown.
 }
\begin{ruledtabular}
\begin{tabular}{|cccccccccc|}
System& $\sqrt{s }$ & collision& rapidity  & b &$\varepsilon_0$ & $\chi^2/N$
& $Q(\chi^2/N,N)$ &$\frac{<N>^{ex}}{<N>^{TH}}$ & $\frac{<p_T>^{ex}}{<p_T>^{TH}}$ \\  
 & (GeV)& Centrality & acceptance & (fm) & $(GeV/fm^3)$ &  & &   \\ \hline
 
Pb+Pb &6.3  &0-7.2 \%&0-1.8 &2.71& $1.32$ &0.69 &0.98 &$2.39\pm 1.43$& $1.22\pm 0.09$ \\
Pb+Pb &7.6  &0-7.2 \%&0-1.8 &2.71& $1.22 $ &3.97 &0.55 &$2.00\pm 0.60$& $1.73\pm 0.12$\\
Pb+Pb &8.8  &0-7.2 \%&0-1.5 &2.71& $1.77 $ &4.52 &0.92 &$1.83\pm 0.33$& $1.33\pm 0.03$\\
Pb+Pb &12.3 &0-7.2 \%&0-1.7 &2.71& $2.87 $ &5.08 &0.75  &$1.61\pm 0.26$& $1.32\pm 0.04$ \\
Pb+Pb &17.3 &0-5.0 \%&0-1.0 &2.3 & $7.15  $ &9.23 & 0.51 &$1.09\pm 0.05$& $1.29\pm 0.03$\\
Au+Au &62.0 &0-20.0\%&-0.5-0.5 &4.43& $18.16 $ &1.52 &0.99 &$1.17\pm 0.22$& $0.93\pm 0.06$\\
Au+Au &130.0&0-11.0\%&-0.5-0.5 &3.2 & $21.62 $ &13.94 &0.12 &$1.50\pm 0.21$& $1.06\pm 0.06$\\
Au+Au &200.0&0-5.0 \%&-0.5-0.5 &2.3 & $31.2 $ &1.02 &0.99 &$1.03\pm 0.09$& $0.91\pm 0.06$
\end{tabular}
\end{ruledtabular}
\end{table*}

\section{Results}\label{sec3}

Details of the data analysed are given in table.\ref{table1}.
NA49 collaboration measured transverse momentum spectra ($\frac{dN}{dydp_T}$) of $\phi$ mesons in
$\sqrt{s }$=6.3, 7.6, 8.8, 12.3 and 17.3 GeV Pb+Pb collisions. 
The measured spectra are shown in five panels (a)-(e) of Fig.\ref{F2}.
In $\sqrt{s }$=6.3-12.3 GeV Pb+Pb collisions,   spectra are measured in 0-7.2\% centrality collisions, while in $\sqrt{s }$=17.3 GeV Pb+Pb collisions centrality cut was 0-5\%. 
STAR measurements for $\phi$ mesons spectra  in $\sqrt{s }$=62, 130 and 200 GeV Au+Au collisions are shown in panels (f), (g) and (h).   Centrality cuts for 
$\sqrt{s }$=62, 130 and 200  GeV Au+Au collisions are 0-20\%, 0-11\% and 0-5\% respectively and the rapidity acceptance is $|y| < 0.5$ .
 
Before we proceed further, it is important to mention that in SPS energy Pb+Pb collisions, the assumption of boost-invariance is not valid. NA49 collaboration \cite{Alt:2008iv} measured rapidity distribution of $\phi$ mesons. In the energy range, $\sqrt{s }$=6.3-17.3 GeV, the rapidity distribution do not show plateau like structure, rather the distribution could be fitted with a Gaussian, width of the Gaussian increasing with energy \cite{Alt:2008iv}. 3+1 dimensional hydrodynamical model is then required to analyze NA49 data. 
The boost-invariant hydrodynamics is used in the spirit of local density approximation. The underlying assumption is that around each rapidity $y$, a narrow window exist where $dN/dy$ is approximately constant.    As will be shown later, with progressive energy, the assumption become more and more accurate.

The NA49 and STAR measurements for the $\phi$ meson $p_T$ spectra are fitted in the model by
varying the central energy density $\varepsilon_0$.  
In Fig.\ref{F2}, the solid   lines  are the best fit obtained to the data. 
 In table.\ref{table1}, best fitted values of the central energy density and corresponding $\chi^2/N$,  for the systems analysed, are noted. In the $\chi^2$ calculations, we have included the statistical and systematic errors \cite{note2}. To test the goodness of fit, in table.\ref{table1}, we have noted the probability $Q(\chi^2/N,N)$,

\begin{equation}
Q((\chi^2/N)_{min},N)=\int_{x=(\chi^2/N)_{min}}^{x=\infty} \chi^2(x,N)dx,
\end{equation}   

\noindent that a random sampling from a Normal distribution would exceed $(\chi^2/N)_{min}$. 
A statistically significant fit require $Q \geq 0.5$.
With the exception of $\phi$ meson data in $\sqrt{s }$=130 GeV Au+Au collisions (quality of which is poor), all the data sets satisfy the requirement. Hydrodynamic model, with the lattice+HRG EOS, reasonably well explain $\phi$ meson spectra over a wide range of energy, $\sqrt{s }$=6-200 GeV.

In table.\ref{table1}, the ratio of the experimental  $\phi$ meson's peak multiplicity over the predicted multiplicity ($\frac{\langle N \rangle^{ex}}{\langle N\rangle^{TH}}$)   , as well as the ratio of the experimental mean $p_T$ over the predicted mean $p_T$ ($\frac{\langle p_T \rangle^{ex}}{\langle p_T\rangle^{TH}}$), are shown. The errors in 
$\frac{\langle N \rangle^{ex}}{\langle N\rangle^{TH}}$ and 
$\frac{\langle p_T \rangle^{ex}}{\langle p_T\rangle^{TH}}$ are experimental only. 
Disagreement with model predictions for $\phi$ multiplicity and experiments increases  as the collision energy is lowered (with the exception of 130 GeV Au+Au data). 
Continual increase of discrepancy between the model predictions and experiments
with lowering energy indicate that the assumption of boost-invariance is violated more and more at lower energy. In the energy range $\sqrt{s }$=6-12 GeV Pb+Pb collisions, NA49 measured $\phi$ mesons in the rapidity range $\Delta y =0-1.5(1.8)$. We have fitted the $p_T$ spectra integrated over the rapidity range.
Over this rapidity range, $\phi$ meson multiplicity changes considerably. 
Naturally, evolution of the fluid, which fit (rapidity) integrated $p_T$ spectra,  underpredict multiplicity at central rapidity. 
$\phi$ mesons mean $p_T$ at SPS energy is also underpredicted in the model.

If beyond a threshold energy, there is a cross over transition from the confined hadronic matter to the deconfined QGP, there will be  a rapid increase in the degeneracy of the medium  (e.g. see Fig.\ref{F1}). Will there be a signature of the transition in the energy range $\sqrt{s }$=6-200 GeV scanned by the $\phi$ mesons? Ratio of the entropy density over cube of the temperature is a measure of the degeneracy of the medium.  $\phi$ meson's multiplicity is expected to be proportional to the initial entropy density. The ratio of $\phi$ multiplicity over the cube of the initial temperature of the fluid is then proportional to the degeneracy of the medium. In Fig.3, collision energy dependence of the ratio of the experimental  $\phi$ meson multiplicity over the cube of the spatially 'averaged' initial temperature $<N>/[.5\cdot N_{part}\cdot \Delta Y \cdot T^3]$ is shown. Multiplicities are normalised by the participant numbers ($N_{part}$) and rapidity gap ($\Delta Y$), to account for the system size, different centrality bins and
different phase space in Au+Au and Pb+Pb collisions. The ratio is approximately constant at RHIC energy $\sqrt{s}$=62-200 GeV. The ratio is approximately constant also at the low SPS energy, $\sqrt{s} \leq$ 12.3 GeV. From low SPS energy to RHIC energy, the ratio increases by a factor of $\sim$ 4. It may be noted (see Fig.\ref{F1}) that in the lattice+HRG EOS, entropy over cube of the temperature also increases by a similar factor ($\sim$ 4.5)    in the temperature range 150-400 MeV. 

In Fig.\ref{F3}, the solid line is a fit obtained to the ratio with an analytical form for the step function,
 $\alpha [1+tanh\frac{\sqrt{s}-\sqrt{s}_{th} }{\Delta \sqrt{s}_{th}}]$. Fitted 
values are, 
$\alpha=1.05\pm 0.21$, $\sqrt{s}_{th}$=$13.65\pm 3.06$ GeV and $\Delta \sqrt{s}_{th}=11.28\pm 5.93$ GeV. The threshold energy $\sqrt{s}_{th}$ can be determined within $\sim$ 20\% accuracy. 
Presently no experimental data exists in the energy range 17.3-62 GeV, between the top of the SPS energy and bottom of the RHIC energy. Uncertainty in the threshold energy could be reduced if the gap is filled. Future STAR energy scan program at RHIC may help to  narrow down the threshold energy  
\cite{Caines:2009yu,Odyniec:2008zz}. It is also  heartening to find the threshold energy close to the top of the SPS energy, where first evidence of QGP formation was obtained \cite{Heinz:2000bk}. The transition is also rather sharp, over a narrow energy range $\Delta\sqrt{s}_{th} \approx$ 12 GeV, the confinement-deconfinement transition occur. 
If the gap between $\sqrt{s}$=17.3-62 GeV is filled in future RHIC energy scan programme, width of the transition may be further reduced.

In the present analysis  we have assumed a baryon free fluid and used the lattice based $\mu_B$=0 equation of state. While at RHIC energy collisions, $\mu_B$=0 is approximately valid, at SPS energy collisions, $\mu_B \neq 0$.  
In a finite baryon density fluid, the confinement-deconfinement transition temperature decreases \cite{Engels:1987rz,de Forcrand:2003hx,deForcrand:2006pv}. Also there is a possibility that at some $\mu_B$, the confinement-deconfinement transition become a 1st order phase transition, with a critical point at some intermediate the baryon density. At the critical point, conserved charges (e.g. baryon density) will have large fluctuations \cite{Cheng:2008zh}. The effect of finite $\mu_B$ on $\phi$ meson production will be an interesting study, which we could not do presently. However, following comments can be made. Other conditions remaining unchanged, in a finite $\mu_B$ fluid,
$\phi$ yield will decrease as the cross-over temperature decreases. Then to reproduce the experimental multiplicity,  initial temperature of $\mu_B \neq 0$ fluid will be higher than $\mu_B=0$ fluid. As an example, we have fitted 
$\phi$ meson $p_T$ spectra in $\sqrt{s}$=6.3 GeV Pb+Pb collisions  with cross-over transition at $T_{co}$=160 MeV. The best fit to the data is obtained with central energy density  $\varepsilon_0$=1.4 $GeV/fm^3$, $\sim$ 6\% higher than the value required for   $T_{co}$=196 MeV.
The fit is shown as the blue line  in Fig.\ref{F2}a. It can not be distinguished from the fit obtained when $T_{co}$=196 MeV. With reduced initial temperature in SPS energy collisions, the ratio $<N>/T_i^3$ will decreases, and the confinement-deconfinement transition will be sharper than estimated presently.   

\section{Summary and conclusions} \label{sec4} 
 
To summarise,  assuming that in high energy heavy ion collisions,  a fluid medium is created, which  become amenable to hydrodynamic description after $\tau_i$=0.6 fm, we have analysed the recent NA49 data on $\phi$ meson production in Pb+Pb collisions in the energy range $\sqrt{s }$=6.3-17.3 GeV and the STAR data on $\phi$ meson production in $\sqrt{s }$=62,130 and 200 GeV Au+Au collisions. 
The hydrodynamic evolution is governed by a lattice+HRG EOS, with a confinement-deconfinement cross-over transition at $T_{co}$=196 MeV. While the confined phase of the EOS is modelled by a
non-interacting hadronic resonance gas, the deconfined part is modelled after a recent lattice simulation. $\phi$ meson production in nuclear collisions, over the energy range $\sqrt{s}$=6.2-200 GeV, are reasonably well explained in the model. We have also tried to extract the threshold energy above which a confined medium crosses over to a
deconfined medium. 
Energy scan by the $\phi$ meson indicate that above a threshold energy $\sqrt{s}_{th}=11.28\pm 5.93$ GeV, the medium crosses over from a confined to a deconfined phase.


\begin{thebibliography}{99}
\bibitem{lattice} 
Karsch F, Laermann E, Petreczky P, Stickan S and Wetzorke I, 
2001 {\it Proccedings of NIC Symposium} (Ed. H. Rollnik and D. Wolf, John 
von Neumann Institute for Computing, J\"ulich, NIC Series, vol.9, 
ISBN 3-00-009055-X, pp.173-82,2002.)

\bibitem{Cheng:2007jq}
  M.~Cheng {\it et al.},
  Phys.\ Rev.\  D {\bf 77}, 014511 (2008)
  [arXiv:0710.0354 [hep-lat]].

\bibitem{BRAHMSwhitepaper}
 BRAHMS Collaboration, I. Arsene {\it et al.},  
Nucl. Phys. A {\bf 757}, 1 (2005). 
 
\bibitem{PHOBOSwhitepaper} 
PHOBOS Collaboration,  B. B. Back {\it et al.},  
Nucl. Phys. A {\bf 757}, 28 (2005). 
 
\bibitem{PHENIXwhitepaper} 
PHENIX Collaboration, K.~Adcox {\it et al.}, 
Nucl. Phys. A {\bf 757} 184 (2005).  
  
\bibitem{STARwhitepaper} 
STAR Collaboration, J. Adams {\it et al.}, 
Nucl. Phys. A {\bf 757} 102 (2005).

  
\bibitem{Koch:1986ud}
  P.~Koch, B.~Muller and J.~Rafelski,
  Phys.\ Rept.\  {\bf 142}, 167 (1986).

\bibitem{Rafelski:1980gk}
  J.~Rafelski and M.~Danos,
  Phys.\ Lett.\  B {\bf 97}, 279 (1980).
  
\bibitem{Cleymans:1998yb}
  J.~Cleymans, H.~Oeschler and K.~Redlich,
  Phys.\ Rev.\  C {\bf 59}, 1663 (1999)
  [arXiv:nucl-th/9809027].

\bibitem{Hamieh:2000tk}
  S.~Hamieh, K.~Redlich and A.~Tounsi,
  Phys.\ Lett.\  B {\bf 486}, 61 (2000)
  [arXiv:hep-ph/0006024].
\bibitem{Tounsi:2002nd}
  A.~Tounsi, A.~Mischke and K.~Redlich,
  Nucl.\ Phys.\  A {\bf 715}, 565 (2003)
  [arXiv:hep-ph/0209284].
  
\bibitem{Becattini:2003wp}
  F.~Becattini, M.~Gazdzicki, A.~Keranen, J.~Manninen and R.~Stock,
  Phys.\ Rev.\  C {\bf 69}, 024905 (2004)
  [arXiv:hep-ph/0310049].
\bibitem{Becattini:2005xt}
  F.~Becattini, J.~Manninen and M.~Gazdzicki,
  Phys.\ Rev.\  C {\bf 73}, 044905 (2006)
  [arXiv:hep-ph/0511092].
 
  
\bibitem{Abelev:2007rw}
  B.~I.~Abelev {\it et al.}  [STAR Collaboration],
  Phys.\ Rev.\ Lett.\  {\bf 99}, 112301 (2007)
  [arXiv:nucl-ex/0703033].
  
 
\bibitem{:2008fd}
  B.~I.~Abelev {\it et al.}  [STAR Collaboration],
  arXiv:0809.4737 [nucl-ex].
\bibitem{Abelev:2008zk}
  B.~I.~Abelev {\it et al.}  [STAR Collaboration],
  Phys.\ Lett.\  B {\bf 673}, 183 (2009)
  [arXiv:0810.4979 [nucl-ex]].

  
\bibitem{Hwa:2006vb}
  R.~C.~Hwa and C.~B.~Yang,
  arXiv:nucl-th/0602024.
  
\bibitem{Alt:2008iv}
  C.~Alt {\it et al.}  [NA49 collaboration],
  Phys.\ Rev.\  C {\bf 78}, 044907 (2008)
  [arXiv:0806.1937 [nucl-ex]].
  
  
\bibitem{note1} We have checked that with the lattice+HRG EOS, in ideal fluid dynamics, STAR measurements of $\frac{dN^\phi}{dy}$ and $<p_T^\phi>$ in 0-5\% Au+Au collisions are best explained with $T_F$=150 MeV.    
 
  
\bibitem{QGP3}
P.~F. Kolb and U. Heinz,
in {\it Quark-Gluon Plasma 3}, edited by R.~C. Hwa and 
X.-N. Wang (World Scientific, Singapore, 2004), p.~634.

\bibitem{Huovinen:2005gy}
  P.~Huovinen,
  Nucl.\ Phys.\  A {\bf 761}, 296 (2005)
  [arXiv:nucl-th/0505036].

\bibitem{note2}
For Au+Au collisions, the systematic error is   $\sim$15\%  \cite{:2008fd}, while that in   Pb+Pb collisions   are approximately same as the statistical ones \cite{Alt:2008iv}.


\bibitem{Caines:2009yu}
  H.~Caines  [STAR Collaboration],
  arXiv:0906.0305 [nucl-ex].


\bibitem{Odyniec:2008zz}
  G.~Odyniec  [STAR Collaboration],
  J.\ Phys.\ G {\bf 35}, 104164 (2008).
  
\bibitem{Heinz:2000bk}
  U.~W.~Heinz and M.~Jacob,
  arXiv:nucl-th/0002042.

\bibitem{Engels:1987rz}
  J.~Engels,
  Nucl.\ Phys.\  A {\bf 461}, 317C (1987).

\bibitem{de Forcrand:2003hx}
  P.~de Forcrand and O.~Philipsen,
  Nucl.\ Phys.\  B {\bf 673}, 170 (2003)
  [arXiv:hep-lat/0307020].
  
\bibitem{deForcrand:2006pv}
  P.~de Forcrand and O.~Philipsen,
  JHEP {\bf 0701}, 077 (2007)
  [arXiv:hep-lat/0607017].

\bibitem{Cheng:2008zh}
  M.~Cheng {\it et al.},
  Phys.\ Rev.\  D {\bf 79}, 074505 (2009)
  [arXiv:0811.1006 [hep-lat]].

 \end{thebibliography}
\end{document}